\documentclass[12pt]{book}
\usepackage{plenum}
\usepackage{graphics}
\usepackage{epsf}

\begin{document}

\chapter{CENTER DOMINANCE, CENTER VORTICES, AND CONFINEMENT}

\author{L. Del Debbio,\refnote{1} M. Faber,\refnote{2} 
J. Greensite,\refnote{3} and {\v S}. Olejn\'{\i}k\refnote{4}}

\affiliation{\affnote{1}CPT, CNRS Luminy, Case 907, 13288 Marseille
Cedex 9, France \\
\affnote{2}Inst. f{\"u}r Kernphysik, Tech. Univ.
Wien, 1040 Vienna, Austria \\
\affnote{3}The Niels Bohr Institute, DK-2100 Copenhagen \O, Denmark \\
\affnote{4}Inst. of Phys., Slovak Acad. of Sci., 
842 28 Bratislava, Slovakia }

\bigskip
\bigskip
\bigskip

{\sl  Invited talk, presented by J. Greensite, at the Nato Workshop 
``New Developments in Quantum Field Theory,'' June 14-20, 1997, 
Zakopane, Poland }

\section{INTRODUCTION}

   The results that I would like to discuss here
are a collection of numerical data
which strongly favor an old and, in recent years, somewhat neglected 
theory of quark confinement: the $Z_N$ Vortex Condensation Theory.
Some of this data (Figs. 1-8) was reported by our group late last
year,\refnote{\cite{Us}} the rest is new.

   The confinement region of an $SU(N)$ gauge theory really consists
of at least two parts.  The first is an intermediate distance region, 
extending from the onset of the linear potential
up to some color-screening distance, which we call
the {\bf Casimir-Scaling regime}.\refnote{\cite{lat96,Cas1}}
Many numerical experiments have
shown that in this intermediate region flux tubes form, and a linear
potential is established, between heavy quarks in any non-trivial
representation of the gauge group.  The string-tension is
representation-dependent, and appears to be roughly proportional to
the quadratic Casimir of the representation.\refnote{\cite{Cas2}}  Thus, for 
an $SU(2)$ gauge theory,
\begin{equation}
        \sigma_j \approx {4\over 3}j(j+1) \sigma_{1/2} 
\end{equation}
where $\sigma_j$ is the string tension for a heavy quark-antiquark pair in 
representation $j$.  Eventually, however, the color charge of 
higher-representation
quarks must be screened by gluons, and the asymptotic string tension
can then only depend on the transformation properties of the quarks under
the center of the gauge group, i.e. on the "n-ality" of the representation.
This {\bf Asymptotic regime} extends from the color-screening 
length to infinity, and in the case of an $SU(2)$ gauge group the 
string tensions must satisfy
\begin{equation}
        \sigma_j = \left\{ \begin{array}{cl}
                        \sigma_{1/2} & j=\mbox{half-integer} \\
                            0    & j=\mbox{integer} \end{array} \right.
\end{equation}
In particular, the string between quarks in an adjoint representation
must break, at some distance which presumably depends on the mass of 
"gluelumps" (i.e. the energy of a gluon bound to a massive adjoint quark).
Also, since string-breaking  is a $1/N^2$ suppressed process, the number of 
colors is relevant.  The breaking of the adjoint string is difficult to 
observe in numerical experiments, although on general theoretical grounds 
one may be confident that the breaking {\it must} occur for sufficiently 
large adjoint quark separation.

   The most popular theory of quark confinement is the abelian projection
theory proposed by 't Hooft, which I will briefly describe in the next
section.  In past years our group has been highly critical of this theory
(as well as the $Z_N$ vortex theory), on the grounds that it fails to
explain the existence of a linear potential between higher representation
quarks in the Casimir scaling regime.\refnote{\cite{lat96,Cas1}}    
This failure is very significant, because it is in the Casimir regime that
the confining force replaces Coulombic behavior, and in fact it is only
in this regime that the QCD string has been well studied numerically.
If we don't understand Casimir scaling, then we don't really understand how
flux tubes form. 

   A possible response to this criticism is simply to admit 
that the formation of flux tubes, at intermediate distances, remains to be 
understood, but that the abelian projection theory is nonetheless valid at 
very large distance scales, i.e. in the asymptotic regime.  I will argue that 
there may be some truth to this response, but that the confining configurations
relevant to the asymptotic regime seem to be $Z_N$ vortices, rather than 
abelian monopoles.

\section{ABELIAN DOMINANCE}

   One of the earliest ideas about confinement, known as 
``dual-superconductivity,'' was put forward independently by 't Hooft
and Mandelstam in the mid-1970's.  The idea is that the QCD vacuum
resembles a superconductor with the roles of the $\vec{E}$ and $\vec{B}$
fields interchanged.  Electric (rather than magnetic) fields are squeezed
into vortices; electric (rather than magnetic) charges are confined.
Magnetic monopoles are condensed; they play the role of the electrically
charged Cooper pairs.  The problem is to actually identify the magnetic
monopoles of an unbroken non-abelian gauge theory, and to understand
which non-abelian degrees of freedom play the role of electromagnetism.

   A concrete suggestion along these lines was made by 't Hooft in
1981.\refnote{\cite{tHooft1}}  The proposal was to gauge fix
part of the $SU(N)$ symmetry by diagonalizing some operator
tranforming in the adjoint representation of the gauge group.  This
leaves a remnant $U(1)^{N-1}$ gauge symmetry,  with gauge
transformations $g$ of the form
\begin{equation}
       g = \mbox{diag}[e^{i\alpha_1},e^{i\alpha_2},...,e^{i\alpha_N}] ~~~~~~~~
       \sum \alpha_n = 0
\end{equation}
The diagonal components of the vector potential, ${\cal A}^{aa}_\mu$, transform
under the residual symmetry like abelian gauge fields, i.e.
\begin{equation}
      {\cal A}'^{aa}_\mu = {\cal A}^{aa}_\mu + \partial_\mu \alpha^a
\end{equation}
while the off-diagonal components transform like double (abelian) charged
matter fields
\begin{equation}
      {\cal A}'^{ab}_\mu = e^{i(\alpha^a-\alpha^b)} {\cal A}^{ab}_\mu
\end{equation}
This gauge-fixed theory can therefore be regarded as an abelian gauge theory
of ``photons,'' charged matter fields, and magnetic monopoles.  Monopole
condensation confines abelian charged objects, and the abelian electric 
field forms a flux tube.  
     
   On the lattice, one can decompose the link variables $U$ 
\begin{equation}
       U_\mu(x) = W_\mu(x) A_\mu(x)
\end{equation}
into ``abelian'' link variables $A$, transforming under the residual 
symmetry as abelian gauge fields, and ``matter'' fields $W$
\begin{eqnarray}
       A'_\mu(x) &=& g(x) A_\mu(x) g^{-1}(x+\hat{\mu})
\nonumber \\
       W'_\mu(x) &=& g(x) W_\mu(x) g^{-1}(x)
\end{eqnarray}
For $SU(2)$ lattice gauge theory, $A$ is simply the diagonal part of $U$,
rescaled to restore unitarity, i.e.
\begin{equation}
       U = a_0 I + i \vec{a}\cdot \vec{\sigma} ~~~ , ~~~~~~~~~~
       A = { a_0 + i a_3 \sigma^3 \over \sqrt{a_0^2+a_3^2} }
\end{equation}

  Monte Carlo studies of the abelian projection theory began with
the work of Kronfeld et al., \refnote{\cite{Kronfeld}} who introduced
a specific abelian projection gauge, the 
``maximal abelian gauge,''\refnote{\cite{Kronfeld1}}
which has been used in most further studies.  The maximal abelian
gauge is defined as the gauge which maximizes the quantity
\begin{equation}
       \sum_x \sum_\mu \mbox{Tr}[U_\mu(x) \sigma^3 U^\dagger_\mu(x) \sigma^3]
\end{equation}
This requires diagonalizing, at every site, the adjoint representation
operator
\begin{equation}
      X(x) = \sum_\mu [U_\mu(x)\sigma^3 U^\dagger_\mu(x) +
U^\dagger_\mu(x-\hat{\mu}) \sigma^3 U_\mu(x-\hat{\mu})]
\end{equation}
This gauge choice makes the link variables as diagonal as possible,
placing most of the quantum fluctuations in the abelian link variables.
If the abelian projection idea is going to work at all, it ought
to work best in this gauge.  Other proposals (Polyakov-line gauge,
Field-Strength gauge) have not, in fact, been very successful.

   An important development was the finding, by Suzuki and collaborators,
that if we fix to maximal abelian gauge and replace the full link 
variables $U$ with the abelian link variables $A$ (this is often termed
``abelian projection''), and then calculate such quantities as
Creutz ratios, Polyakov lines, etc., with the abelian links,
the results very closely approximate those obtained with the
full link variables.\refnote{\cite{Suzuki}}  The fact that the abelian 
link variables seem
to carry most of the information about the infrared physics is known
as ``{\bf abelian dominance},'' and it
has stimulated a great deal of further work on the abelian 
projection theory.

\section{CENTER DOMINANCE}

   Of course the abelian projection theory is not the {\it only} proposal
for explaining the confining force; there have been many other suggestions
over the years.  One idea that was briefly popular in the late
1970's was the Vortex Condensation theory, put forward, in
various forms, by 't Hooft,\refnote{\cite{tHooft2}} 
Mack,\refnote{\cite{Mack}} and by 
Nielsen and Olesen\refnote{\cite{CopVac}} (the ``Copenhagen
Vacuum'').  The idea is that the QCD vacuum
is filled with closed magnetic vortices, which have the topology
of tubes (in 3 Euclidean dimensions) or surfaces (in 4 dimensions) of 
finite thickness, and which carry magnetic flux in the center of
the gauge group (hence ``center vortices'').  
The effect of creating a center vortex linked to a given Wilson 
loop, in an $SU(N)$ gauge theory, is to multiply the
Wilson loop by an element of the gauge group center, i.e.
\begin{equation}
        W(C) \rightarrow e^{i2\pi n/N} W(C) ~~~~~~~ n=1,...,N-1
\end{equation}
Quantum fluctuations in the number of vortices linked to a Wilson
loop can be shown to lead to an area law falloff, assuming that
center vortex configurations are condensed in the 
vacuum.\footnote{Some related ideas have also been put forward
by Chernodub et al.\refnote{\cite{Poli1}}}

   With one notable exception,\refnote{\cite{Tomboulis}} almost
nothing has been done with this idea since the early 1980's, which
was at the dawn of Monte Carlo lattice gauge simulations.  It is therefore
interesting to go back and study the vortex theory, using
a numerical approach inspired by studies of the abelian 
projection theory.   

   In an $SU(2)$ lattice gauge theory, we begin 
by fixing to maximal abelian gauge and then go one step further,
using the remnant $U(1)$ symmetry to bring the abelian link variables
\begin{equation}
       A = \left[ \begin{array}{cc}
                   e^{i\theta} & \\
                            & e^{-i\theta} \end{array} \right]
\end{equation}
as close as possible to the $SU(2)$ center elements $\pm I$ by
maximizing $<\cos^2 \theta>$, leaving a remnant $Z_2$ symmetry.
This is the {\bf (indirect) Maximal Center Gauge} (the center is
maximized in $A$, rather than directly in $U$).
We then define at each link
\begin{equation}
          Z \equiv \mbox{sign}(\cos\theta) = \pm 1
\end{equation}
which is easily seen to transform like a $Z_2$ gauge field under the
remnant $Z_2$ symmetry.  ``Center Projection'' is the replacement 
$U \rightarrow Z$ of the full link variables by the center variables; we can
then calculate Wilson loops, Creutz ratios, etc. with the center-projected
$Z$-link variables.

   Figure 1 is a plot of Creutz ratios vs. $\beta$, extracted from the
center-projected configurations.  The straight line is the
asymptotic freedom prediction
\begin{equation}
     \sigma a^2 = {\sigma \over \Lambda^2} \left({6\pi^2 \over 11}\beta
\right)^{102/121}
               \exp\left[-{6\pi^2 \over 11}\beta \right]
\end{equation}
with the value $\sqrt{\sigma}/\Lambda = 67$.  What is remarkable about
this plot, apart from the scaling, is that the Creutz ratios $\chi(R,R)$
at each $\beta$ are almost independent of $R$.  This means that the center 
projection sweeps away the Coulombic contribution, and the linear potential 
appears already at $R=2$.  This is seen quite clearly in Fig. 2, which compares
the center-projected Creutz ratios (solid line), at $\beta=2.4$, with
the Creutz ratios of the full theory (dashed line).  It is also interesting
to compute the Creutz ratios derived from abelian link variables with 
the $Z$ variable factored out, i.e. $A/Z$ (dotted line). We note that, in 
this case, the string tension simply disappears.  

\begin{figure}[h!]
\scalebox{.6}{\rotatebox{90}{\includegraphics{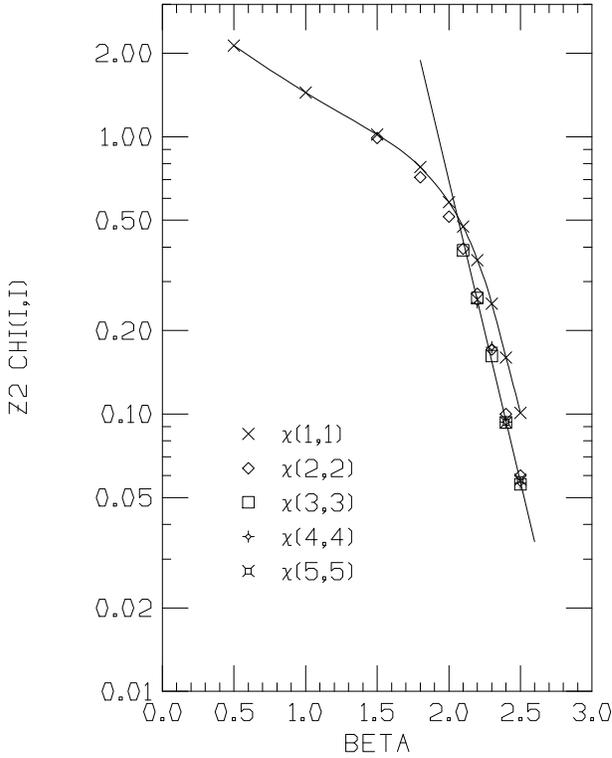}}}
\caption{Creutz ratios from center-projected lattice
configurations, in the (indirect) maximal center gauge.}
\end{figure}

\begin{figure}[h!]
\scalebox{.4}{\rotatebox{90}{\includegraphics{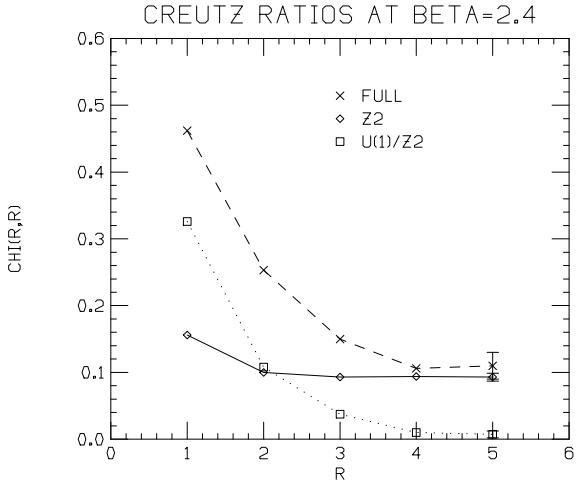}}}
\caption{Creutz ratios $\chi(R,R)$ vs. R at $\beta=2.4$ for
full, center-projected, and $U(1)/Z_2$-projected lattice configurations.}
\end{figure}

   It seems evident from this data that, just as the abelian $A$ links are the
crucial part of the full $U$ link variables in maximal abelian gauge,
so the $Z$ center variables are the crucial part of the $A$ links in
maximal center gauge, carrying most of the information about the
string tension.  This is what we mean by ``Center Dominance.''  

   Should one then interpret center dominance to mean that the confining
force is due to $Z_2$ center vortices, rather than $U(1)$ monopoles?
That conclusion would be premature, in our view.
In fact, our original interpretation of 
this data was that the success of center dominance suggests that {\it neither}
abelian dominance {\it nor} center dominance has anything very convincing to 
say about quark confinement (and this fits very nicely with our further 
critique of abelian projection
based on Casimir scaling).\refnote{\cite{lat96}}  Underlying that
interpretation, however, was the belief that the ``thin'' $Z_2$ vortices of the
center-projected configurations are probably irrelevant to the
confining properties of the full, unprojected configurations.  This belief 
is testable, however, and the result of the test is surprising.

\section{VORTEX-LIMITED WILSON LOOPS}

   The only excitations of $Z_2$ lattice gauge theory with non-zero action
are ``thin'' $Z_2$ vortices, which have the topology of a surface
(one lattice spacing thick) in D=4 dimensions.  We will call the
$Z_2$ vortices, of the center projected $Z$-link configurations, 
``Projection-vortices'' or just {\bf P-vortices}. 
These are to be
distinguished from the hypothetical ``thick'' center vortices, which
might exist in the full, unprojected $U$ configurations.  A plaquette 
is pierced by a P-vortex if, upon going to maximal center gauge and 
center-projecting, the projected plaquette has the value $-1$.
Likewise, a given lattice surface is pierced by $n$ P-vortices if 
$n$ plaquettes of the surface are pierced by P-vortices.

   In a Monte Carlo simulation, the number of P-vortices piercing
the minimal area of a given loop $C$ will, of course, fluctuate.  Let
us define $W_n(C)$ to be the Wilson loop evaluated on a sub-ensemble
of configurations, selected such that precisely $n$ P-vortices, in
the corresponding center-projected configurations, pierce the minimal area
of the loop.  It should be emphasized here that the center projection
is used only to select the data set.  The Wilson loops themselves are
evaluated using the full, {\it unprojected} link variables.  In practice,
to compute $W_n(C)$, the procedure is to generate thermalized lattice 
configurations by the usual Monte Carlo algorithm, and fix to maximal
center gauge by over-relaxation.  For each independent configuration
one then examines each rectangular loop 
on the lattice of a given size; those with $n$ P-vortices piercing the
loop are evaluated, the others are skipped.  Creutz ratios $\chi_n(I,J)$
can then be extracted from the vortex-limited Wilson loops $W_n(C)$.
In particular, if the presence or absence of P-vortices in the 
projected configuration is unrelated to the confining properties of
the corresponding unprojected configuration, then we would expect
\begin{equation}
      \chi_0(I,J) \approx \chi(I,J)
\end{equation}
at least for large loops.

   The result of this test is shown in Fig. 3.  Quite contrary to
our expectations, the confining force vanishes if P-vortices are
excluded.  This does not necessarily mean that the confining
configurations of $SU(2)$ lattice gauge theory are thick center
vortices.  It does imply, however, that the presence or absence
of P-vortices in the projected gauge field is strongly correlated with
the presence or absence of confining configurations (whatever they
may be) in the unprojected gauge field. 

\begin{figure}[h!]
\scalebox{.6}{\includegraphics{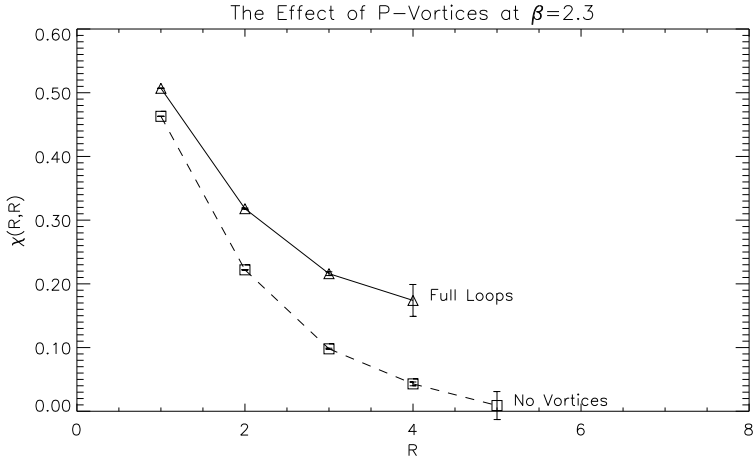}}
\caption{Creutz ratios $\chi_0(R,R)$ extracted from loops with
no P-vortices, as compared to the usual Creutz ratios $\chi(R,R)$,
at $\beta=2.3$.}
\end{figure}

   The next question is whether we can rule out the possibility
that the confining configurations are, in fact, thick $Z_2$ center
vortices.  Suppose, for a moment, that to each P-vortex in the projected
$Z$-link gauge field there corresponds a thick center vortex in the associated,
unprojected, $U$-link gauge field.  If that is the case, then in the limit
of large loop area we expect
\begin{equation}
       {W_n(C) \over W_0(C)} \longrightarrow (-1)^n
\label{WnW0}
\end{equation}
The argument for this equation is as follows:
Vortices are created by discontinuous gauge transformations.  Suppose 
loop $C$, parametrized by $x^\mu(\tau),~\tau \in [0,1]$, 
encircles $n$ vortices.  
At the point of discontinuity
\begin{equation}
       g(x(0)) = (-1)^n g(x(1))
\end{equation}
The corresponding vector potential, in the neighborhood of loop $C$ can
be decomposed as
\begin{equation}
       A^{(n)}_\mu(x) = g^{-1}\delta A^{(n)}_\mu(x) g + i g^{-1} 
\partial_\mu g
\end{equation}
so that
\begin{eqnarray}
       W_n(C) &=& <\mbox{Tr}\exp[i\oint dx^\mu A_\mu^{(n)}]>
\nonumber \\ 
          &=& (-1)^n <\mbox{Tr}\exp[i\oint dx^\mu \delta A_\mu^{(n)}]> 
\end{eqnarray}
In the region of the loop $C$, the vortex background looks locally like
a gauge transformation.  If all other fluctuations $\delta A^{(n)}_\mu$ are
basically short-range, then they should be oblivious, in the neighborhood
of the loop $C$, to the presence or absence of vortices in the middle of
the loop.  In that case, if we have correctly identified the vortex
contribution, then
\begin{equation}
<\mbox{Tr}\exp[i\oint dx^\mu \delta A_\mu^{(n)}> ~ \approx ~
      <\mbox{Tr}\exp[i\oint dx^\mu \delta A_\mu^{(0)}]> 
\end{equation}
for sufficiently large loops, and eq. (\ref{WnW0}) follows
immediately.  All we have to do is test this.
 
   Figures 4 and 5 show our data for $W_1/W_0$ and $W_2/W_0$, respectively,
at $\beta=2.3$.  Again, somewhat to our surprise, this data is entirely
consistent with (\ref{WnW0}); it is consistent with the confining field
configurations being center vortices, and in fact offers good evidence
in favor of that possibility.

\begin{figure}[h!]
\scalebox{.6}{\includegraphics{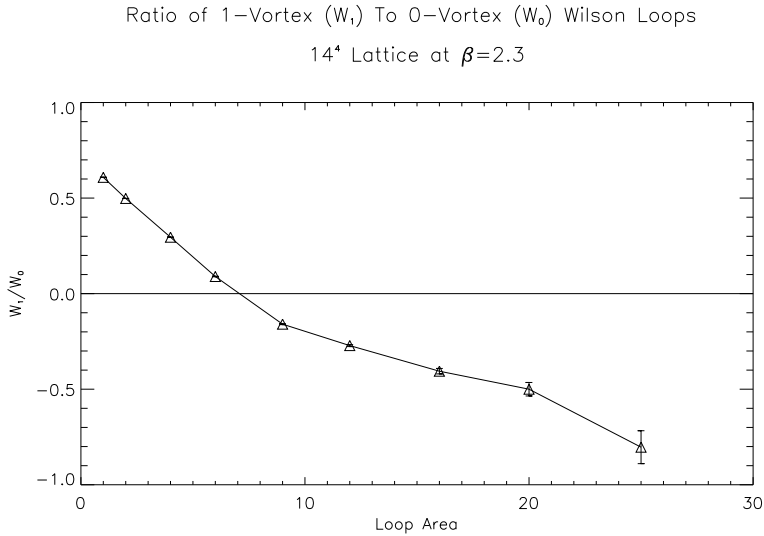}}
\caption{Ratio of the 1-Vortex to the 0-Vortex Wilson loops,
$W_1(C)/W_0(C)$, vs. loop area at $\beta=2.3.$}
\end{figure}

\begin{figure}[h!]
\scalebox{.6}{\includegraphics{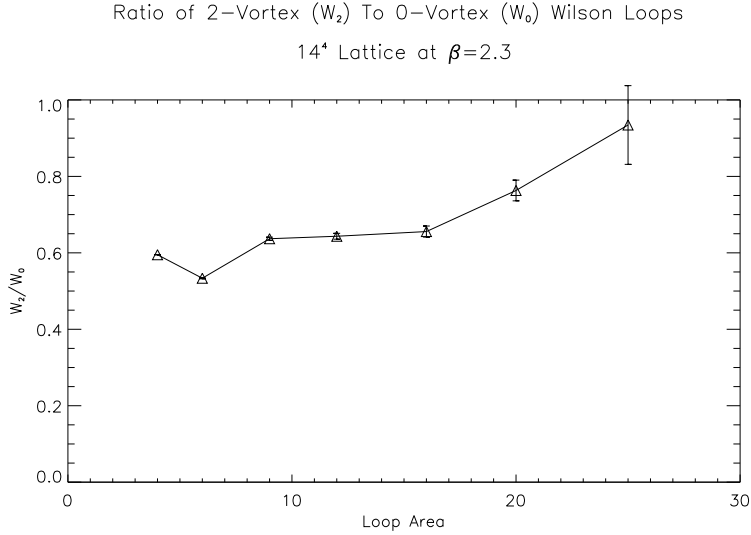}}
\caption{Ratio of the 2-Vortex to the 0-Vortex Wilson loops,
$W_2(C)/W_0(C)$, vs. loop area at $\beta=2.3$.}
\end{figure}

   Of course, it could still be that we are looking at a rather
small (and perhaps misleading) sample of the data, at least for
the larger loops.  Large loops will tend to be pierced by large
numbers of P-vortices.  As the area of a loop increases, the fraction
of configurations in which no P-vortex (or exactly
one, or exactly two P-vortices) pierces the loop
will decrease, tending to zero in
the limit.  So let us instead consider $W_{evn}(C)$ and $W_{odd}(C)$,
where $W_{evn}(C)$ denotes Wilson loops evaluated in
configurations with an even (including zero) number of P-vortices
piercing the loop, and $W_{odd}(C)$ denotes the corresponding quantity for
odd numbers. Then
\begin{equation}
      W(C) = P_{evn}(C) W_{evn}(C) + P_{odd}(C) W_{odd}(C)
\label{evn-odd}
\end{equation}
where
\begin{tabbing}
$P_{evn}(C) = $ \= the fraction of configurations with an even (or zero) \\
                \> number of P-vortices piercing loop $C$ \\ \\
$P_{odd}(C) = $ \> the fraction of configurations with an odd \\
                \> number of P-vortices piercing loop $C$ 
\end{tabbing}
One expects that for large loops, $P_{evn} \approx P_{odd} \approx 0.5$.
According to the vortex condensation mechanism, neither $W_{evn}$ 
nor $W_{odd}$ falls with an area law; the area-law falloff is due to a 
delicate cancellation between the two terms in eq. (\ref{evn-odd}).  
As loops become large, one should find $W_{odd} \rightarrow - W_{evn}$.  The 
data, shown below in Figures 6-8, support these expectations.  This time
we are using essentially all of the data, since about half contributes
to $W_{evn}(C)$, and the rest to $W_{odd}(C)$.

\begin{figure}[h!]
\scalebox{.6}{\includegraphics{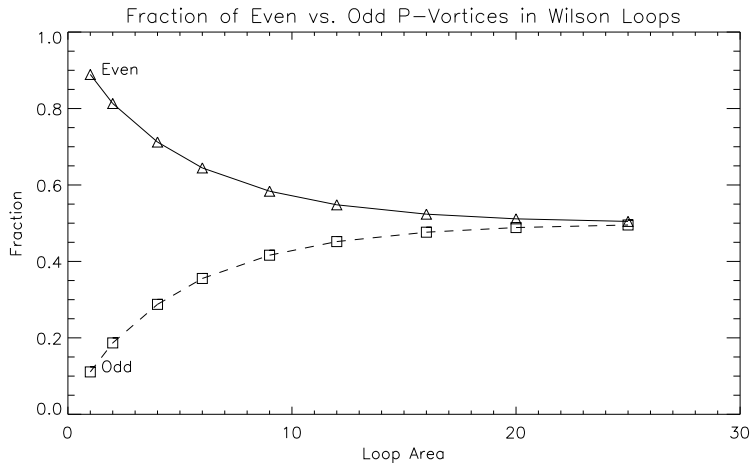}}
\caption{Fraction of link configurations containing 
even/odd numbers of P-vortices, at $\beta=2.3$, piercing loops of
various areas.}
\end{figure}

\begin{figure}[h!]
\scalebox{.6}{\includegraphics{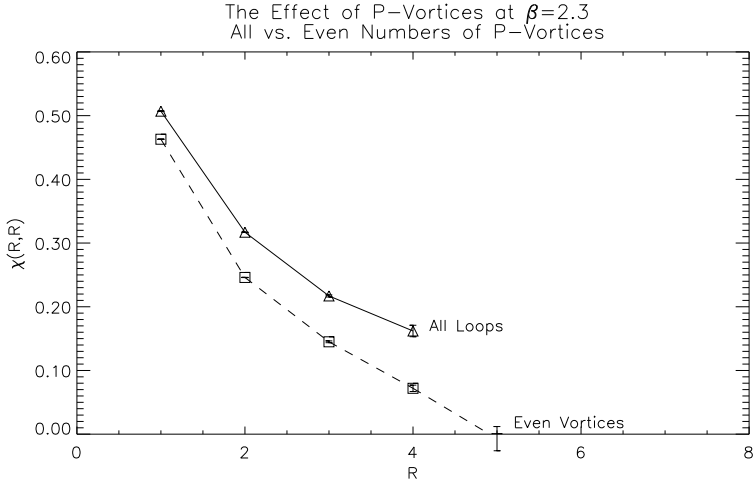}}
\caption{Creutz ratios $\chi_{evn}(R,R)$ extracted from
Wilson loops $W_{evn}(C)$, taken from configurations with
even numbers of P-vortices piercing the loop.  The standard Creutz
ratios $\chi(R,R)$ at this coupling ($\beta=2.3$) are also shown.}
\end{figure}

\begin{figure}[h!]
\scalebox{.6}{\includegraphics{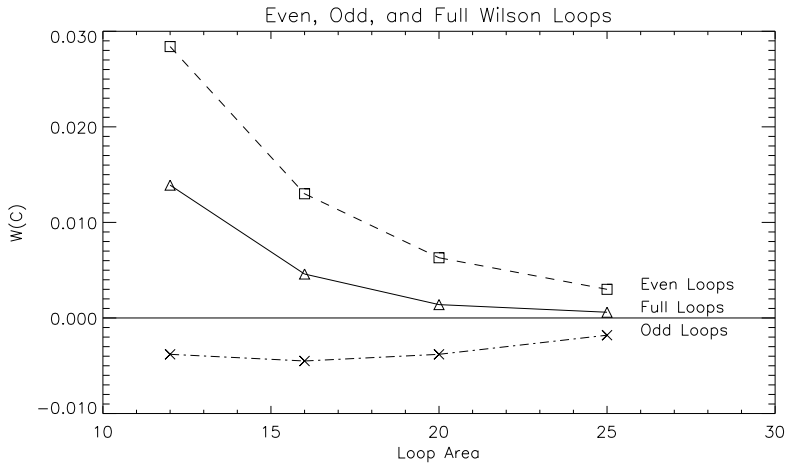}}
\caption{Wilson loops $W_{evn}(C),~W_{odd}(C)$ and $W(C)$ at
larger loops areas, taken from configurations with even numbers of
P-vortices, odd numbers of P-vortices, and any number of P-vortices,
respectively, piercing the loop.  Again $\beta=2.3$.}
\end{figure}

\section{DIRECT MAXIMAL CENTER GAUGE}

   Along with the successes, there is one significant failure
of center dominance in the data shown in Fig. 1.  Despite the
nice scaling of the data, the value of $\sqrt{\sigma}/\Lambda=67$ is
a little high, and in fact the center projected Creutz ratios are all 
significantly higher than the asymptotic string tension extracted
from unprojected configurations, using ``state-of-the-art'' noise
reduction techniques.

   On the other hand, it is not so clear that the ``indirect'' maximal center
gauge is the {\it true} maximal center gauge.  What we have done
up until now is to first fix to maximal abelian gauge, and then
bring the abelian part $A$ of link $U$ as close as possible to
$\pm I$.  However, since we are emphasizing the role of the gauge
group center, rather than the $U(1)$ subgroup, it really makes more 
sense to choose a gauge in which the entire link variable $U$ is
brought as close as possible to the center elements $\pm I$.
With this motivation, let us define the
{\bf (direct) Maximal Center Gauge} of an $SU(N)$ 
gauge theory as the gauge which maximizes the quantity
\begin{equation}
        Q = \sum_x \sum_\mu \mbox{Tr}[U_\mu(x)]\mbox{Tr}[U_\mu^{\dagger}(x)]
\end{equation}
For the $SU(2)$ gauge group, we define
\begin{equation}
        Z = \mbox{sign(Tr$[U]$)}
\end{equation}
as the center-projected link variables; these again transform like
$Z_2$ gauge fields under the remnant $Z_2$ gauge symmetry.

   Using the direct maximal center gauge, we find the following results:
Qualitatively, things look about the same,
and plots of $W_n/W_0$, and $W_{evn}$ vs. $W_{odd}$, look virtually identical 
to the previous data in the indirect maximal center gauge.
Quantitatively, however, there is an improvement.  We find that
string tensions extracted from the
center projection in the ``direct'' gauge are in much better agreement with the
asymptotic string tension of the full theory, extracted by 
``state-of-the-art'' methods. 
Figure 9 shows a plot of Creutz ratios vs. $\beta$. The straight line
is the usual scaling curve, but this time with a value
$\sqrt{\sigma}/\Lambda=58$.  Figures 10-12 plot the center-projected
Creutz ratios $\chi(R,R)$ at $\beta=2.3,~2.4,~2.5$ respectively.  The triangles
are our data.  The solid line is the asymptotic string tension of
the unprojected configurations at these values of $\beta$, quoted by
Bali et al.\refnote{\cite{Bali}} The dashed lines are the error bars on the 
asymptotic string tension, which we have also taken from this reference.

\begin{figure}[h!]
\scalebox{.6}{\rotatebox{90}{\includegraphics{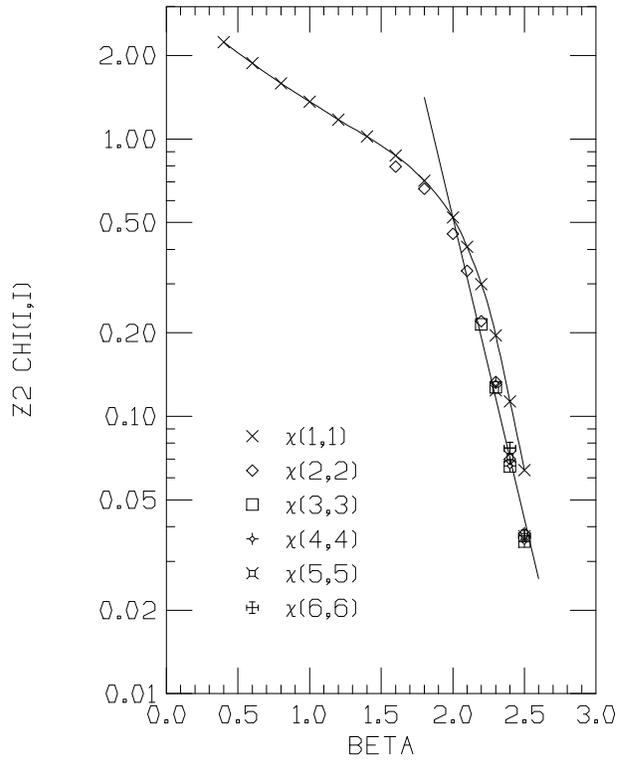}}}
\caption{Creutz ratios from center-projected lattice configurations,
in the direct maximal center gauge.}
\end{figure}

\begin{figure}[h!]
\scalebox{.6}{\includegraphics{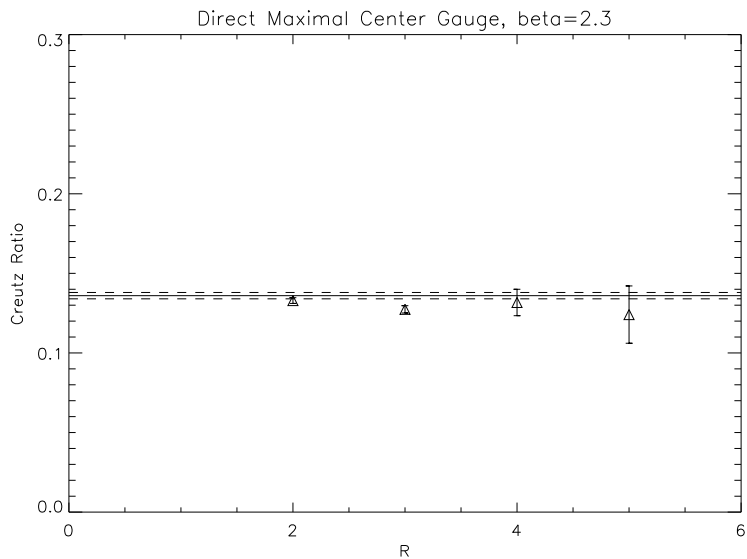}}
\caption{Center-projection Creutz ratios $\chi(R,R)$ vs. $R$ at
$\beta=2.3$; direct maximal center gauge.  Triangles are our data
points. The solid line shows the value of the asympotic string tension
of the unprojected configurations, and the dashed lines the associated
error bars, quoted in Bali et al.\refnote{\protect\cite{Bali}} }
\end{figure}

\begin{figure}[h!]
\scalebox{.6}{\includegraphics{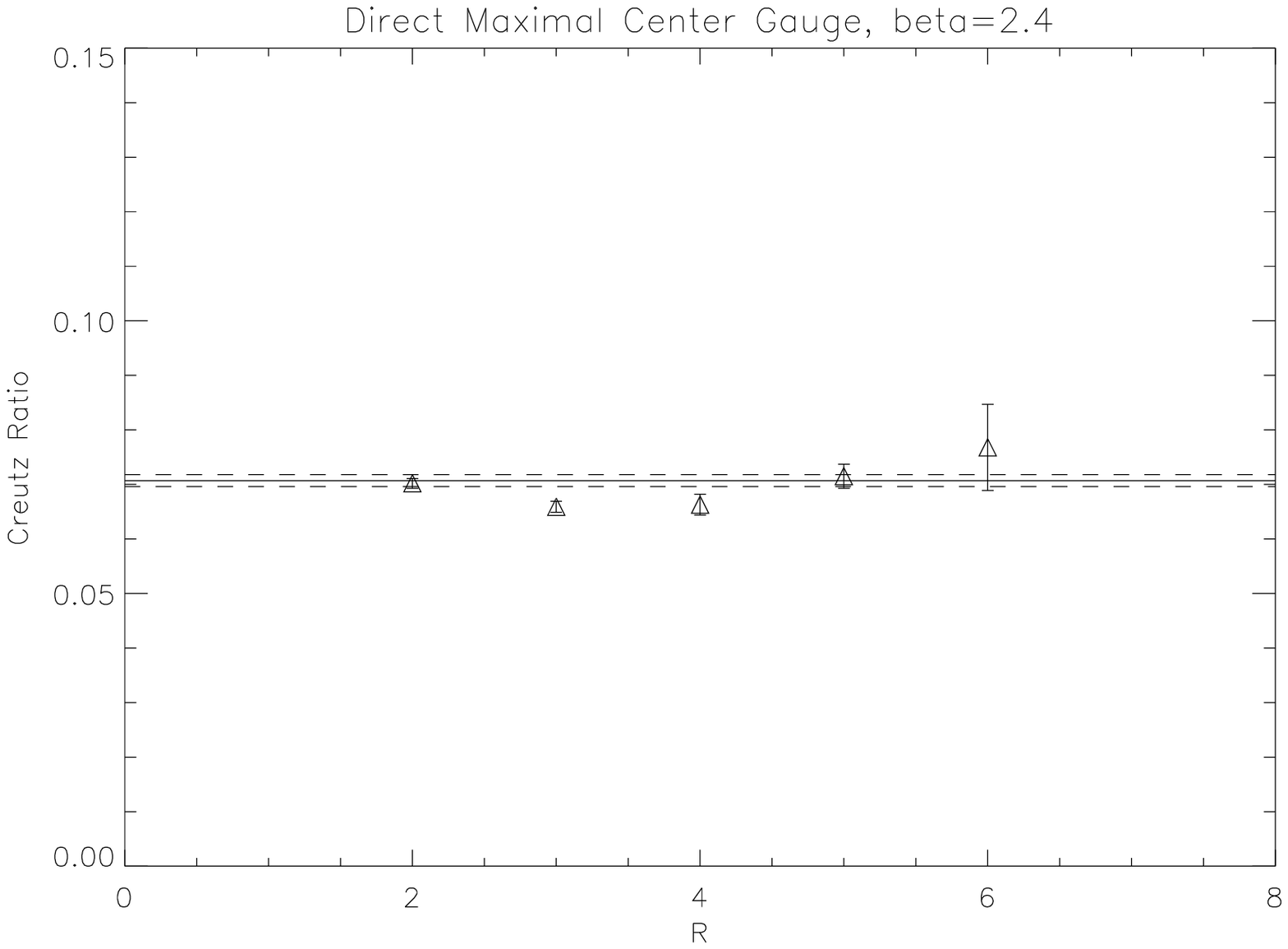}}
\caption{Same as Fig. 10, at $\beta=2.4$.}
\end{figure}

\begin{figure}[h!]
\scalebox{.6}{\includegraphics{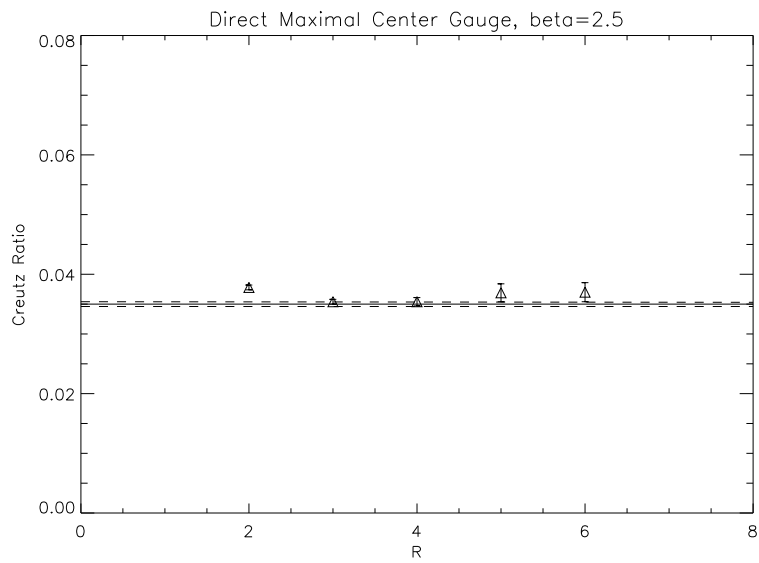}}
\caption{Same as Fig. 10, at $\beta=2.5$.}
\end{figure}

\clearpage

\section{VORTICES VS. MONOPOLES}

   There is no denying that the data shown here, in support of
the vortex condensation theory, is a little reminiscent of the data
that has been put forward in support of the abelian projection theory.
This raises a natural question:  If the Yang-Mills vacuum is dominated,
at long wavelengths, by $Z_2$ vortex configurations, then how do we
explain the numerical successes of the abelian projection in maximal
abelian gauge?  In our opinion, the probable answer to this question is 
that a center vortex configuration, transformed to maximal abelian gauge
and then abelian-projected, will appear as a chain of monopoles
alternating with antimonopoles.  These monopoles are essentially
an artifact of the projection; they are condensed because the long
vortices from which they emerge are condensed.

  A little more graphically, the picture is as follows:  Consider a
center vortex at some constant time $t$.  This time-slice of a
thick vortex is then a tube
of magnetic flux.  Before gauge-fixing, the field-strength inside this
tube points in arbitrary directions in color space


\begin{figure}[h]
\scalebox{.8}{\includegraphics{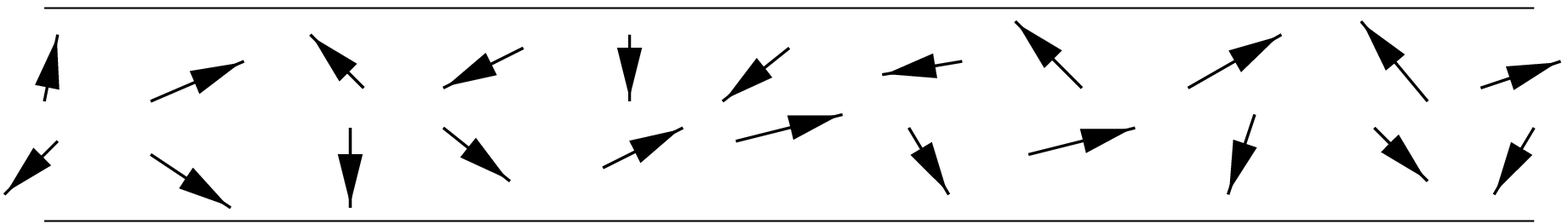}}
\end{figure}


\noindent Fixing to maximal abelian gauge, the field strength tends to line up
mainly (but not entirely) in the diagonal ($\pm \sigma^3$) color direction


\begin{figure}[h]
\scalebox{.8}{\includegraphics{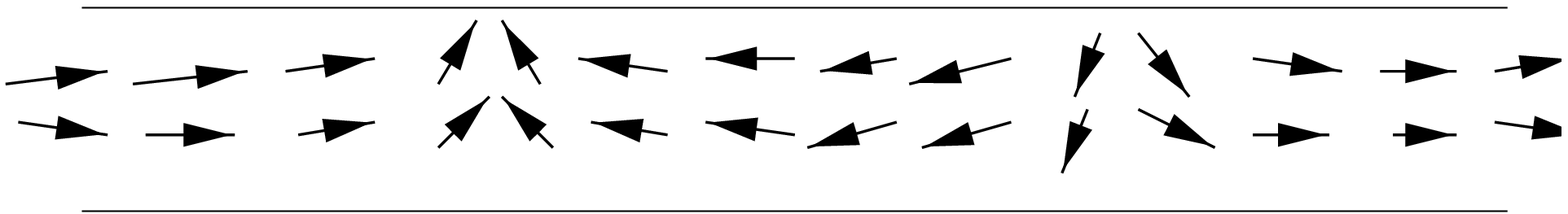}}
\end{figure}


\noindent Upon abelian projection, the regions 
interpolating between $+\sigma^3$
and $-\sigma^3$ emerge as ``monopoles.''  Their location is gauge
(and Gribov copy) dependent.


\begin{figure}[h]
\scalebox{.8}{\includegraphics{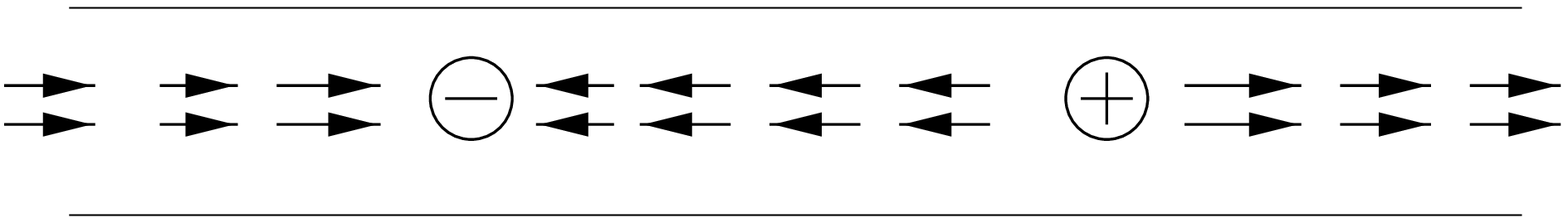}}
\end{figure}


\noindent It is not difficult to construct examples of center vortices which
behave in just this way, i.e. which are converted to monopole-antimonopole
chains upon abelian projection in maximal abelian gauge.  
  
\newpage

   If this picture is accurate, then the ``spaghetti vacuum''
  
\vspace{32pt}

\begin{figure}[h]
\scalebox{.90}{\includegraphics{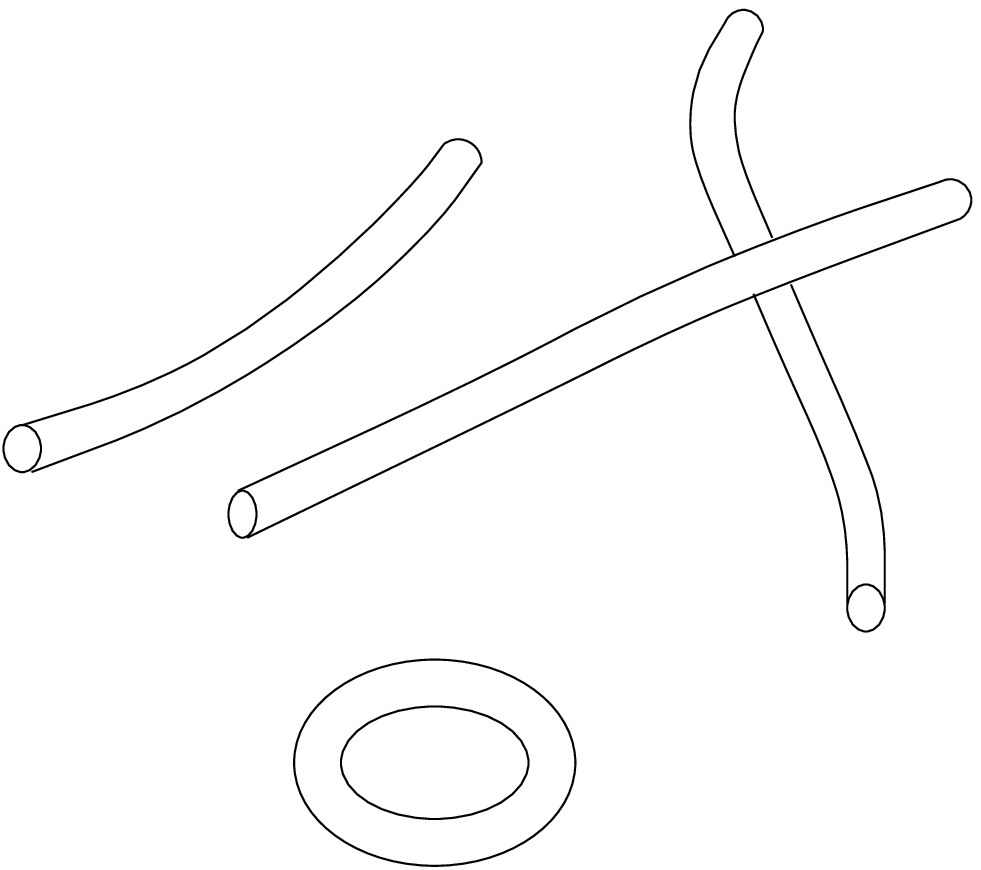}}
\end{figure}

\vspace{32pt}

\noindent appears, under abelian projection, as a ``monopole vacuum''

\vspace{32pt}

\begin{figure}[h]
\scalebox{.90}{\includegraphics{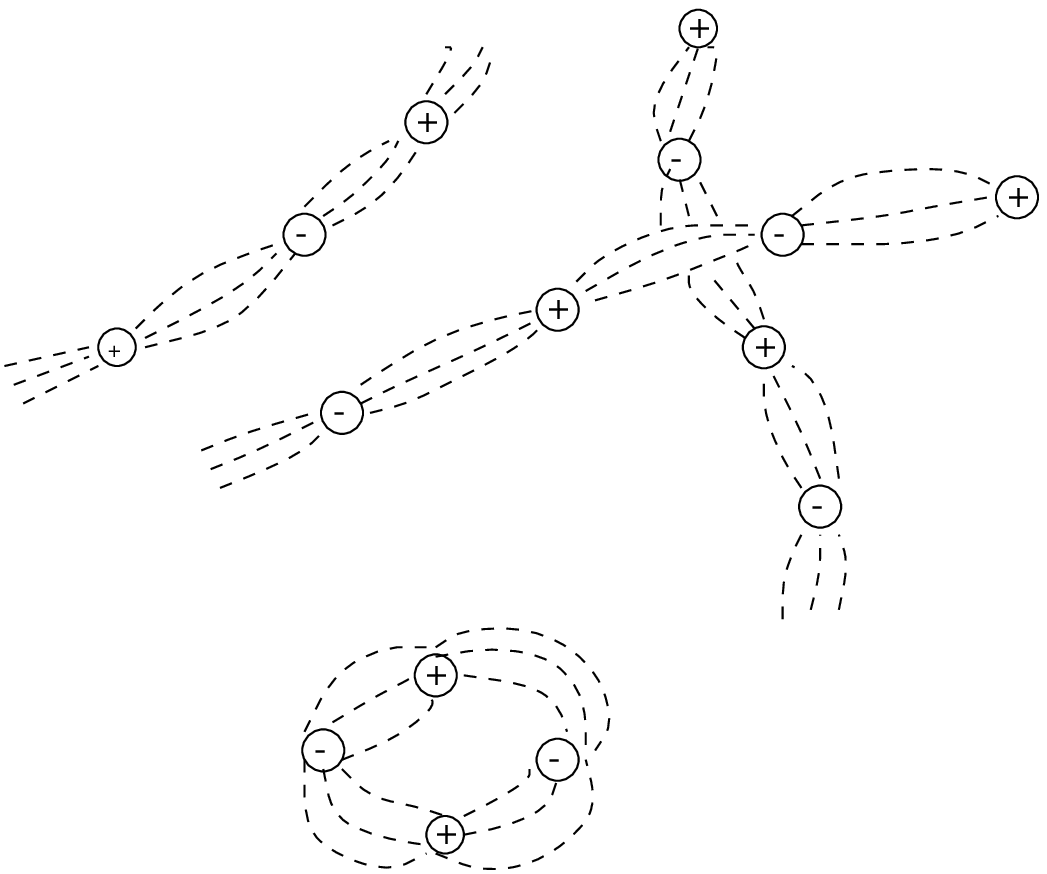}}
\end{figure}

\clearpage

   We have, in fact, obtained  some preliminary evidence for this 
picture from Monte Carlo simulations.  These simulations were
carried out at $\beta=2.4$, in the (indirect) maximal center gauge.
We look at sites where the monopoles are ``static,''
i.e. the monopole current is $j_0=\pm 1,~\vec{j}=0$.
The monopole charge is enclosed in a cube bounded by spacelike 
plaquettes.  We find that:

\vspace{30pt}

\begin{description}

\item{I:}  Almost all ($93\%$) of monopole cubes are pierced by one,
and only one, P-vortex.

\bigskip

\begin{figure}[h]
\centerline{\epsfbox{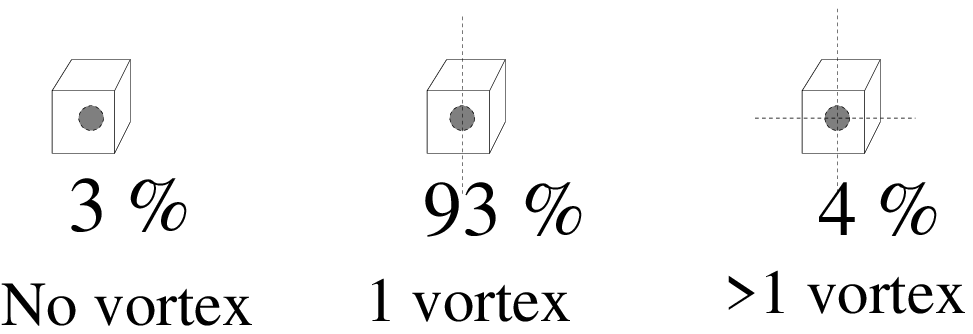}}
\end{figure}

\vspace{40pt}

\item{II:} The action of a monopole cube, pierced by a P-vortex, 
is highly asymmetric.  Almost all the plaquette action
\begin{equation}
       S = (1 - {1\over 2} \mbox{Tr}[UUU^{\dagger}U^{\dagger}]) - S_0
\end{equation}
above the lattice average $S_0$, is oriented in the
direction of the P-vortex.  On each of the two plaquettes pierced by 
the P-vortex, at $\beta=2.4$, the average action above $S_0$ is $S=0.29$.
On each of the four plaquettes which are not pierced by the vortex,
$S=0.03$ on average.\footnote{Bakker et al.\refnote{\cite{Poli2}} 
have also studied the
excess action of monopole cubes (but not the correlation with 
P-vortices) in maximal abelian gauge.}

\bigskip

\begin{figure}[h]
\centerline{\epsfbox{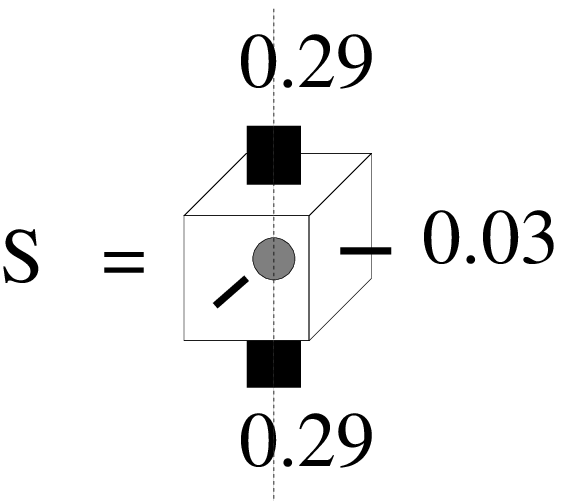}}
\end{figure}

\newpage

\item{III:} The (unprojected) action distribution of a monopole cube,
pierced by a P-vortex, is similar to the action distribution of
any other cube pierced by a P-vortex...

\bigskip
\bigskip

\begin{figure}[h]
\centerline{\epsfbox{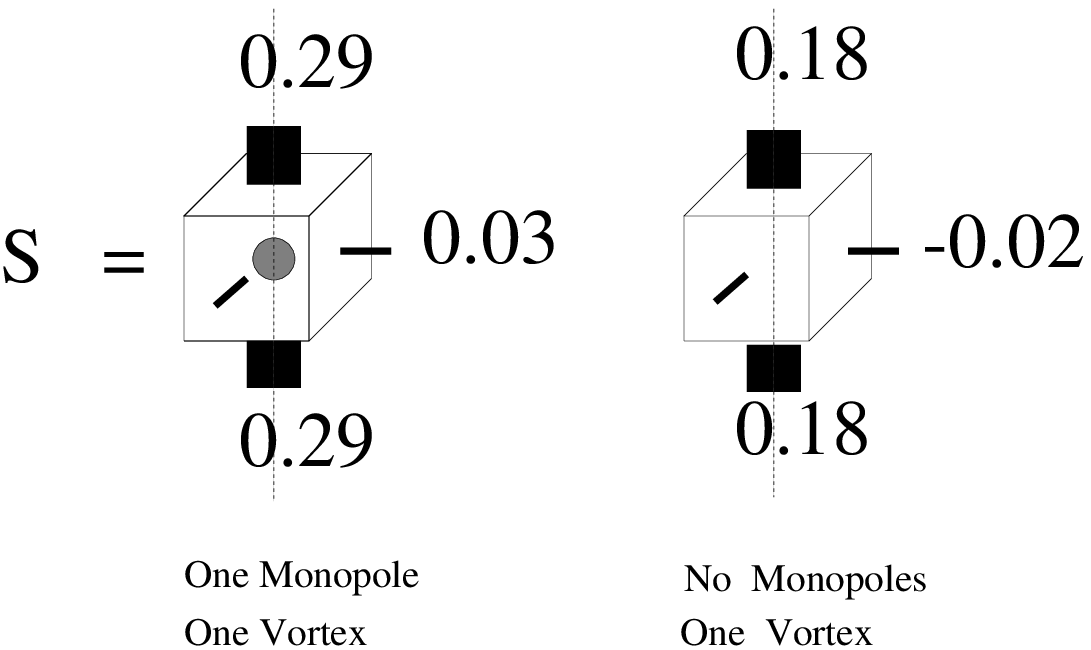}}
\end{figure}

\bigskip

...especially when we look at ``isolated'' monopoles (no neighboring
monopole currents) 

\bigskip
\bigskip

\begin{figure}[h]
\centerline{\epsfbox{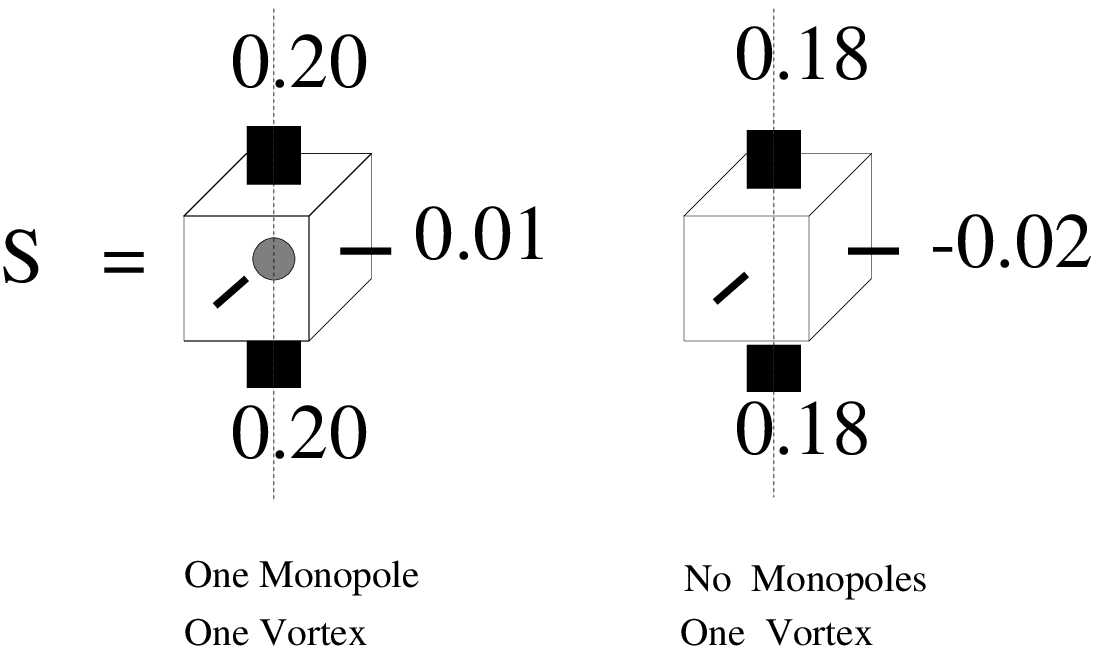}}
\end{figure}

\bigskip

\end{description}

\noindent In summary, abelian monopoles tend to lie along P-vortices.
Isolated monopoles are hardly distinguished, in their (unprojected) field
strength distribution, from other regions along the P-vortices.
\bigskip

   This is in accordance with our intuitive picture.

\clearpage

\section{CONCLUSIONS}

   We have developed a technique for locating center vortices in
thermalized lattice gauge configurations, and have found evidence 
that center vortices account for
the asymptotic string tension between static, fundamental
representation, color charges.  A ``spaghetti vacuum'' picture appears
to be correct at sufficiently large scales.

   On the other hand, string formation at intermediate distances, in the
Casimir scaling regime, remains to be understood.  This is a very important
issue, especially since the Casimir scaling regime extends to infinity as
$N_{colors} \rightarrow \infty$.\refnote{\cite{JGH}}  
Casimir scaling suggests that 
center vortices, although they may be the crucial configurations 
asymptotically, are not the whole story.  Since adjoint loops are oblivious
to the gauge-group center,  
one may speculate that there are other types of configurations which 
contribute to the adjoint string tension.  Or, possibly, the 
finite thickness
and detailed inner structure of center vortices is a relevant issue, since 
adjoint loops which intersect the ``core'' of a center vortex {\it will} 
be affected by the vortex.  Perhaps the 
gluon-chain model,\refnote{\cite{gluon}} which I
proposed some time ago, might be helpful in understanding the
dynamics of the Casimir-scaling region.

   We are currently in the process of repeating all our calculations
for $SU(3)$ lattice gauge theory, and have already found evidence 
of center dominance on small lattices at strong couplings.
If we also find that (i) center dominance persists on larger lattices 
at weaker couplings; (ii) the absence of P-vortices results in vanishing
string tension, and (iii) 
\begin{equation}
        {W_n(C) \over W_0(C)} ~ \longrightarrow ~ e^{2n\pi i/3}
\end{equation}
then the combined evidence in favor of some version of the $Z_N$ vortex 
condensation theory will be quite compelling.


  One final note: Shortly after the Zakopane meeting, Tomboulis and
Kov\'{a}cs reported on some new Monte Carlo data they have obtained
in support of the vortex condensation theory.\refnote{\cite{TK}}
Their results are quite consistent with the work I have presented here.

\section{ACKNOWLEDGEMENTS}

   I would like to thank the organizers of the ``New Developments''
meeting for inviting me to 
Zakopane.  I am also grateful for the hospitality of the 
high-energy theory group at Lawrence Berkeley National Laboratory, 
where some of this work was carried out.  This research was supported
in part by the U.S. Department of Energy, under Grant No. DE-FG03-92ER40711.
Support has also been provided by Carlsbergfondet.

\end{document}